%% file: ms.tex
\documentclass[12pt,preprint]{aastex}

\slugcomment{}

\shorttitle{NGC 2516}

\shortauthors{Terndrup et al.}

\newcommand{\angs}{\AA\ }         % with a space

\newcommand{\kms}{km s$^{-1}$}    % with no space

\newcommand{\kmss}{km s$^{-1}$\ } % with a space

\newcommand{\secmark}{\S\ }       % ApJ section symbol

\newcommand{\vsini}{$v \sin i$}   % v sin i no space

\newcommand{\vsinis}{$v \sin i$\ }% with a space

\begin{document}

\title{Rotation and activity in the solar-metallicity open
cluster NGC~2516\altaffilmark{1,2}}

\author{Donald M. Terndrup\altaffilmark{3} and Marc
Pinsonneault} \affil{Department of Astronomy, Ohio
State University, 140 West 18th Avenue, Columbus,
OH 43210} \email{terndrup@astronomy.ohio-state.edu,
pinsono@astronomy.ohio-state.edu}

\author{Robin D. Jeffries and Alison Ford\altaffilmark{4}}
\affil{Department of Physics, Keele University,
Keele, Staffordshire ST5 5BG, United Kingdom}
\email{rdj@astro.keele.ac.uk, Alison.Ford@open.ac.uk}

\author{John R. Stauffer\altaffilmark{3}} \affil{Infrared
Processing and Analysis Center, California Institute of
Technology, Mail Code 100-22, 770 South Wilson Avenue,
Pasadena, CA 91125} \email{stauffer@ipac.caltech.edu}

\and

\author{Alison Sills} \affil{Department of Physics
and Astronomy, McMaster University, 120 Main
Street West, Hamilton, Ontario L8S 4M1, Canada}
\email{asills@mcmail.cis.mcmaster.ca}

\altaffiltext{1}{Based on observations obtained at the
Anglo-Australian Telescope.}

\altaffiltext{2}{Based on observations obtained at the
Cerro Tololo-Interamerican Observatory, NOAO, which is
operated by the Associated Universities for Research in
Astronomy (AURA), Inc., under cooperative agreement with
the National Science Foundation.}

\altaffiltext{3}{Visiting Astronomer, Cerro Tololo
Inter-American Observatory, NOAO.}

\altaffiltext{4}{Present address, Physics and Astronomy
Group, The Open University, Walton Hall, Milton Keynes
MK7 6AA, United Kingdom.}

\shorttitle{Rotation in NGC~2516}

\shortauthors{Terndrup et al.}

\begin{abstract} We report new measures of radial
velocities and rotation rates ($v \sin i$) for 51 F
and early-G stars in the open cluster NGC~2516, and
combine these with previously published data.  From high
signal-to-noise spectra of two stars, we show that NGC~2516
has a relative iron abundance with respect to the Pleiades
of $\Delta{\rm [Fe/H]} = +0.04 \pm 0.07$ at the canonical
reddening of $E(B - V) = 0.12$, in contrast to previous
photometric studies that placed the cluster 0.2 to 0.4 dex
below solar.  We construct a color-magnitude diagram based
on radial velocity members, and explore the sensitivity
of photometric determinations of the metallicity and
distance to assumed values of the reddening.  For a metal
abundance near solar, the Hipparcos distance to NGC~2516
is probably underestimated.  Finally, we show that
the distribution of rotation rates and X-ray emission
does not differ greatly from that of the Pleiades, when
allowance is made for the somewhat older age of NGC~2516.
\end{abstract}


\section{Introduction}

Observations of stellar rotation rates in young clusters
demonstrate that stars must lose a considerable fraction
of their initial angular momentum both before and after
arrival on the main sequence.  In young clusters such
as Alpha Persei (age $\sim 50$ Myr) or the Pleiades (age
$\sim 100$ Myr), stars at any mass are seen to rotate at
a variety of rates \citep[e.g.,][]{sta89,que98,ter00},
with most rotating slowly ($v \sin i \leq 10 - 15$ \kms),
but a minority rotating much more quickly ($v \sin i
\geq 30$ \kms).  By the age of the Hyades (600 Myr),
all but the lowest-mass stars have spun down to slow
rotation rates \citep{rad87,sta87,jon96,ter00,tpt02}.
This picture is reinforced by observations in clusters
with ages between the Pleiades and the Hyades, such as
M~34 = NGC~1039 \citep{sod01} or NGC~6475 \citep{jj97},
although in these clusters the number of stars with
measured rotation rates is smaller than in the more
well-studied systems.  Here and throughout this paper,
ages come from \citet{mmm93}, unless otherwise stated;
we do not use ages derived from the ``lithium depletion
boundary'' technique \citep{ssk98,bar99,st99}, because
few clusters have ages measured in this manner.

The current evolutionary paradigm for stellar angular
momentum loss has a number of ingredients that
are motivated by observations or theory.  The main
sequence spindown is attributed to magnetic coupling
between the stellar envelope and an ionized wind
\citep[e.g.,][]{sch62,wd67}, which would cause a star to
spin down over tens to hundreds of millions of years.
The existence of rapid rotators in clusters of age $<
200$ Myr requires that the angular momentum loss rate
must be suppressed or saturated at high rotation rates
\citep{kaw88,mb91}.  The presence of slow rotators in young
clusters points to an additional loss mechanism acting over
lifetimes of a few Myr \citep{cc93,kep95}; motivated by
the theoretical work of \citet{kon91} and \citet{shu94},
which showed that magnetic torques can transfer angular
momentum away from a protostar onto its accretion disk,
this early loss is termed ``disk locking.''

Much of the effort to understand the evolution of spin
rates has concentrated on using clusters of various
ages to set constraints on the duration of disk-locking
and mass-dependence of saturation used in the models
\citep{kri97,all98,sil00}.  An important but often
unstated assumption in many of these studies is that open
clusters of different ages sample a single time sequence
of angular momentum loss.  For this to be the case,
the distribution of initial specific angular momenta
and disk locking lifetimes, as well as the properties of
saturation and of winds, would be the same in all open
clusters (specifically, the dependence on stellar mass
would be identical).  Testing this assumption has always
been difficult since most of the available data were
for Alpha Persei, the Pleiades, and the Hyades, which
mainly sample the loss mechanisms on the main sequence.
Consequently there have been a number of efforts underway
to obtain new observations of more clusters of similar age
\citep[e.g.,][]{jfss97,jj97}, or by extending observations
to very young systems such as the Orion Nebula Cluster
\citep{ah92,smmv99,hrhc00,hbm01,reb01}.  Recently,
\citet{tpt02} have shown that a single set of parameters
describing the angular momentum loss can explain the
rotational evolution from Orion to the Pleiades and Hyades.

The southern open cluster NGC~2516 provides an important
additional laboratory for exploring the evolution of
angular momentum in young clusters.  NGC~2516 is a
fairly rich cluster with a mass about twice that of
the Pleiades, in which many possible members have been
identified photometrically \citep{jth01}.  There have
been radial velocity studies of bright stars near the
main-sequence turnoff \citep{gl00}, and a previous study
of the rotation rates of F-G stars \citep{jjt98} which we
extend in this paper.  The cluster has also been the target
of several X-ray surveys \citep{jtp97,mic00,har01,sci01},
and has a distance determination from Hipparcos parallaxes
\citep{rob99}.

NGC~2516 has been of particular interest because
several previous photometric studies have suggested
a low metallicity, a few tenths dex below solar
\citep{cam85,jtp97,jjt98,pin98}.  In contrast, a
preliminary metallicity estimate from an equivalent
width analysis of relatively low signal-to noise spectra
\citep{jjt98} indicated an abundance for NGC~2516 that
was solar within the errors.  As in any other cluster,
the cluster age from the main-sequence turnoff depends
sensitively on the metallicity.  If NGC~2516 is near solar
metallicity, it has an age of 140 Myr, but would have an
age about 180 Myr for [Fe/H] $\approx -0.3$, a difference
of 30\%.  Furthermore, accurate age determinations enter
into many other aspects of stellar evolution in open
clusters, such as the relation between the initial and
final masses of white dwarfs and the maximum mass of white
dwarf progenitors \citep{kr96,rdj97,vhg00,wei00}.

Determining the metallicity of NGC~2516 is also critical
for using the cluster as a probe of the metallicity
dependence of angular momentum loss.  Theory shows
that, once on the main sequence, the depth of the
outer convective zone depends mainly on stellar mass,
somewhat on metallicity, and hardly at all on stellar
age \citep[e.g.,][]{pdc01}.  Stars of low metallicity
have shallower convective zones at a given effective
temperature, and because low-metallicity stars of a
given mass are also hotter and less luminous than their
solar-abundance counterparts, their convection zones should
be thinner than higher-abundance stars of the same color.

Here we present the results of a spectroscopic study of
NGC~2516, which combines parallel efforts led by two of us
(DMT and RDJ).  In \secmark 2, we discuss the observations
and data reduction, and proceed in \secmark 3 to derive
a metallicity for NGC~2516 both from new spectra and from
analysis of the color-magnitude diagram.  In \secmark 4,
we discuss the rotational properties of the stars in our
combined sample, and present our conclusions in \secmark 5.

\section{Observations and Data Reduction}

\subsection{Photometry}

Photometry comes from a new catalog prepared by
\citet{jth01} from a CCD survey of 0.86 square degree
of NGC 2516, obtained at the CTIO 0.9-m telescope during
three nights in January 1995. Conditions were photometric
and the individual CCD frames were flat-fielded with
twilight exposures. Nightly corrections for extinction and
transformations to the Johnson $V$, $B - V$ and Cousins $V
- I_{\rm C}$ systems were achieved using many observations
of standard fields from the \citet{lan92} compilation.
The numbers of standards observed ensures that, for stars
of the colors considered in this paper, statistical errors
in transforming to the standard colors and magnitudes
can be neglected.

Aperture photometry was performed using small apertures
(either 3 or 6 arcsec radius) on a series of 25 mosaicked
CCD fields, each covering an area of 13.5$\times$13.5
arcminutes$^{2}$ at a scale of 0.4 arcsec/pixel.
Comparison of stars in the overlapping regions of the CCD
fields gives us a good idea of the internal photometry
errors.  For the stars considered in this paper, with $V =
11 - 15$, these internal errors are 0.022 mag in $V$, and
0.015 mag in $B - V$ and $V - I_{\rm C}$.  Residuals to
the linear transformation functions used to fit the Landolt
standards were less than 0.02 mag on all nights.

\subsection{Spectroscopy at Cerro Tololo}

Most of our new spectra were obtained in two runs at
the Blanco 4~m telescope at Cerro Tololo in March 1998
and January 1999.  For both runs, we used the echelle
spectrograph with ``red long'' camera to produce spectra
with an average dispersion of $\approx 0.08$ \angs
pixel$^{-1}$.  The effective resolution as measured from
the FWHM of arc lamp lines was 2.4 pixels, or 8.7 \kmss
at H$\alpha$.  The spectra taken in 1998 used the G181 cross
disperser (316 l mm$^{-1})$, and covered the wavelength
interval 5840 -- 8150 \angs in 28 orders.  In 1999,
we used the 226-3 cross disperser (226 l mm$^{-1}$) to
obtain spectra from 5180 \angs to 8280 \angs in 39 orders.
The wavelength coverage in each order exceeded the free
spectral range, so the wavelengths at the ends of each
order were sampled twice.  In each run, we obtained spectra
of bright field stars with known radial velocity spanning
the temperature range of our NGC~2516 targets, and took
spectra of the evening twilight sky.
Most of the spectra from CTIO have signal-to-noise of
15-25 per resolution element, but we obtained repeated long
exposures of two stars for an abundance analysis of
NGC~2156, below.

Basic image processing and spectral reductions were
performed with scripts written for the {\sc IRAF}\footnote
{IRAF is distributed by the National Optical Astronomy
Observatory, operated by AURA, Inc., under cooperative
agreement with the National Science Foundation.} and
{\sc VISTA} packages \citep{sto88}.  After removal of the
overscan level and a zero-exposure frame, each image was
processed with a moving-box median filtering algorithm
to remove the many cosmic rays.  The spectral orders were
then traced using the exposures of bright stars, and the
sky and object apertures were defined for each object to
maximize the number of pixels contributing to the sky.
Finally, the spectra were extracted using an unweighted sum
over the spectral aperture after background subtraction.
This background subtraction also removed scattered light
from the spectra.

The spectrograph at CTIO exhibits considerable flexure with
telescope position,  amounting to $\pm 1.5$ pixels at the
declination of NGC 2516 during each night.  We corrected
for this by cross correlating selected orders of the
extracted spectra containing the A and B atmospheric bands,
and then applied the measured pixel shifts to bring each
spectrum to a uniform zero point.  The accuracy of this
process was estimated by the scatter in the derived
pixel shifts for the four orders that had significant
telluric absorption.  The average error was typically 0.1
pixel, which is equivalent to 0.36 \kmss at H$\alpha$,
but was occasionally higher.  This error was later added
in quadrature to the error in the radial velocity.

Wavelength calibration was achieved using ThAr lamp spectra
in two steps.  In the first, lines were identified and a
trial solution was obtained independently for each order.
In the second step, quadratic polynomials were fit to the
central wavelength and dispersion as a function of order
number, and the line identifications and solutions for the
wavelength at each pixel were rederived using this fit as
a starting point. In this step, as before, each order had
an independent wavelength solution.

Radial velocities were measured by cross correlation
of seven extracted orders of each spectra against the
equivalent orders of bright, high signal-to-noise template
stars in the lists of \citet{udry99} and \cite{slt99}.
The seven orders were chosen to have many stellar
absorption features and to be relatively free of telluric
absorption, and had central wavelengths from 6040 \angs
to 6778 \AA.  Each order was separately cross correlated
and the results averaged.  The scatter in the velocities
for the separate orders was used to estimate the error
in the velocity.  The lowest velocity errors, about 0.5
\kms, were found for the most narrow-lined stars (i.e.,
the most slowly rotating ones) and/or those stars with
higher signal to noise.  For stars of moderate to rapid
rotation, the radial velocity error is typically about
1/20 the value of \vsini.  The total error was taken as
the quadrature sum of the radial velocity error and the
error in the flexure correction.  The average error for
the entire sample was about 1.1 \kms.

The zero point of the radial velocity system was defined
by the stars HR~2593 (spectral type K2), HR~4540 (F8),
HR~4695 (K1), and HR~5353 (G5).  The estimated external
error for the velocities in this system is 0.4 \kms.

The width of the cross correlation between each star and
the template was used to determine the rotational velocity.
The calibration of the relation between the correlation
widths and \vsinis was determined by convolving the
spectrum of another high signal-to-noise spectrum of a
slowly-rotating star with rotational broadening functions,
then cross-correlating those artificially broadened spectra
with the template star spectrum \citep[cf.][]{har86}.
The broadening functions are given by \citet{gray92};
we adopted a limb-darkening coefficient $\epsilon = 0.6$.

\subsection{Spectroscopy at the AAT}

Data were taken on the AAT on 2000 February 23-24.
The UCL Echelle Spectrograph with 79 lines mm$^{-1}$
grating was used at the Coud\'{e} focus, with a MITLL3
chip. The central wavelength was 6150{\AA} in order 37.
A 1.2$\arcsec$ slit was used, giving a resolution of
0.13{\AA} at 0.036{\AA} pixel$^{-1}$  (in order 33).

The data were reduced at the Keele Starlink node, using
the {\sc echomop}, {\sc figaro} and {\sc uclsyn} packages
\citep{mil97,smi92}.  As with the CTIO data, the spectra
reduction here included background subtraction (including
scattered light), and wavelength calibration. 

Radial velocities were measured using cross-correlation
techniques in the spectral range 5210-5340{\AA} and
5460-5600{\AA}, which contain many neutral metal lines. The
standard templates used to determine heliocentric RVs
were radial velocity standards HR 1829 (G5 V) and HR 4540
(F9 V). Internal errors are dominated by small shifts in
the wavelength calibrations during exposures amounting to
about 1 \kms. External errors, found by comparing standard
star observations, are estimated to be 0.4 \kms.

Rotational velocities were obtained by fitting a synthetic
spectrum to the data, allowing the line abundances and
rotational velocity to vary until the minimum $\chi^2$
value was obtained. The atmospheres used were those of
\citet{kur93} and used a mixing length approximation
with no overshooting.  Oscillator strengths were fixed
by comparison with a solar spectrum taken with the same
instrumentation.

The spectral range used was 5260-5290{\AA}, which was in
the order with the deepest lines.  Temperatures for the
synthesis were calculated from the $B - V$ photometry given
in the catalog of \citet{jth01}, using the prescription
of \citet{sh85}, which includes a dependence on [Fe/H].
For this analysis, we assumed [Fe/H] $= -0.32$, but the
results would be nearly the same had we assumed solar
abundance.\footnote{The metallicity of the atmosphere
used in the analysis has little effect on the line widths,
but would change the relation between color and effective
temperature.  Adopting solar metallicity would have made
the models $\approx 100$ K hotter.  The difference in the
intrinsic width of the lines for this temperature change
does not produce a significant effect in the derived value
of \vsini.} Errors in \vsinis were obtained by finding the
\vsinis at which the $\chi^2$ values increased by an amount
appropriate for the number of degrees of freedom. This
was added in quadrature to an uncertainty of around 0.7
\kms, which arises from uncertainties in the effective
temperature put into the synthesis (a change in $B - V$
of 0.02 changes the temperature by 75~K at 6500~K).

\subsection{Results and internal checks}

Table 1 presents our measurements in NGC~2516.  Column 1
lists the name of each star, where DK indicates the list of
\citet{dk89}, E indicates \citet{egg72}, and the remaining
names are from the photometric catalog of \citet{jth01} or
those used in \citet{jjt98}.  Columns 2 -- 6 display the
source for the data, the radial velocity and its error,
and the rotational velocity and its error.  Finally, the
last five columns show the photometry, catalog entry,
and a photometric membership flag for each star from
\citet{jth01}.  The first part of the table shows stars
which we class as cluster members because their radial
velocities are statistically consistent with the cluster
mean velocity, while the second part shows stars with
discrepant radial velocities.  These memberships were
assigned by an iterative scheme that determined which
stars were within $3\sigma$ of the cluster mean velocity.
We determined the mean by combining our observations
here with the velocities for members in \citet{jjt98}.
The reported errors were increased by adding 1 \kmss in
quadrature to account for the likely velocity dispersion
in the cluster \citep{st97}.

Six of the stars in the CTIO sample had previously been
observed by \citet{jjt98}.  The weighted mean difference in
radial velocity in the sense (CTIO -- previous) is $+1.1$
kms, indicating that the CTIO data have a slightly higher
zero point in velocity than the earlier work.  The weighted
scatter in the difference is 0.9 \kms, where a value of 1.3
\kmss would be the expected error from the random 
errors in the velocities.  Table 2 lists the mean
cluster velocity from this study and earlier works;
the error in that table includes only random errors.
The mean velocity for our full sample is slightly higher
than derived in \citet{jjt98}, but is not statistically
significant given the external errors here and that paper
($\approx 0.4$ \kms).  Our cluster velocity is
distinctly greater than the value of $+19$ \kmss quoted in
the \citet{lyn87} catalog, and is somewhat greater than
the value of $22.0 \pm 0.2$ \kmss derived for early-type
stars and red giants in NGC~2516 by \citet{gl00}. The
latter study included a correction of $-0.6$ \kmss for the
gravitational redshift of main-sequence stars, a correction
we do not apply here.  Had we applied such a correction,
our mean cluster velocity would still be slightly higher
than in \citet{gl00}.

\begin{figure}\epsscale{0.7} \plotone{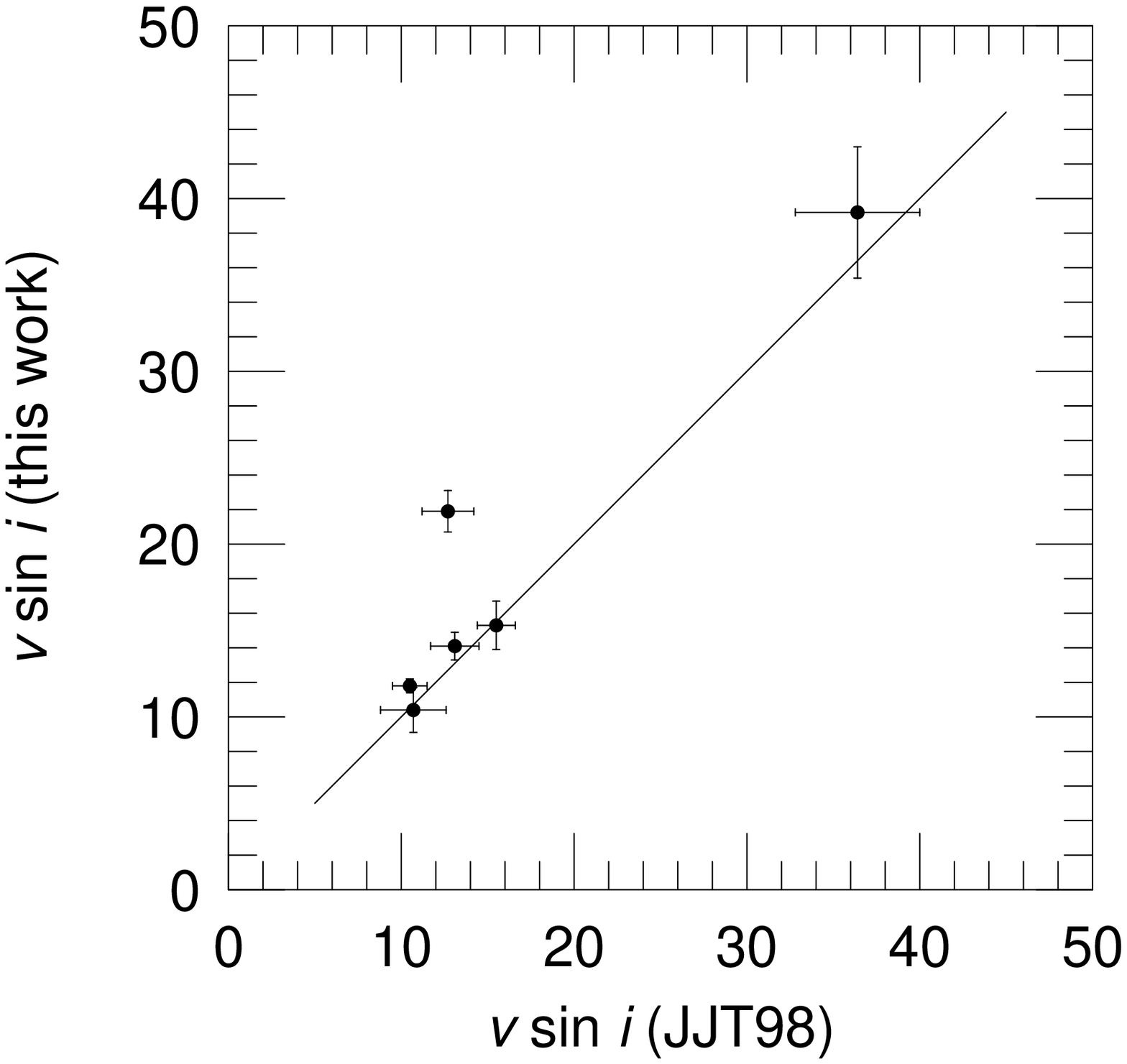}
\caption{Comparison of new and previously published values
of \vsini.  The ordinate displays \vsinis values from this
work, while the abscissa shows those measured by Jeffries
et al.\ (1998).  The $1\sigma$ error bars are also shown.
The discrepant point is for the star DK~206, discussed in
the text.\label{vsinicomp}} \end{figure}

Figure \ref{vsinicomp} shows the correlation of the new
and older values of \vsini;  the solid line on that figure
indicates equality and is not a fit to the data points.
Five of the stars agree within the errors, while one star
(DK~206) has a \vsinis value from CTIO that is considerably
greater than found previously.  We are unable to discern
the source of this discrepancy, although we have verified
that the recorded coordinates for this star on the CTIO
run match the catalog position of DK~206.

All but two of the stars we observed were listed as
probable members in \citet{jth01}, where membership was
determined by proximity to the main sequence in both the
$V$, $B - V$ and $V$, $V - I$ color-magnitude diagrams and
in the $B - V$, $V - I$ two-color plot.  The exceptions
are DK~573 and DK~742, which passed two of the tests.

Of the stars listed as non-members (lower part of Table 1),
most are photometric members in \citet{jth01}.  Three of
these (DK~417, DK~669, and DK~919) have velocities within
20 \kmss of the cluster mean.  Of these, only DK~417 has
a detected rotational velocity (\vsinis $ = 17.9$ \kms),
and so may be a binary member of NGC~2516.  DK~669 is
listed as a photometric binary candidate in \citet{jth01}.
Several other stars have more discrepant velocities but
also exhibit significant rotation (DK~306, DK~503, and
DK~508); these may be candidate short-period binaries.  
Of these three, only DK~508 is a binary candidate from
photometry. Second-epoch radial velocities would be helpful to 
further elucidate their nature.  To be conservative,
we have not included the non-member stars in
the discussion below.

\section{Analysis}

\subsection{The abundance of NGC 2516}

We obtained repeated, long exposures of two slowly-rotating
stars (DK~320 and DK~325); the resulting average spectra
are of sufficient quality (peak S/N of about 70 per
pixel) to attempt a fine spectroscopic abundance analysis.
From the point of view of open cluster studies, the best
approach is to define the abundance relative to other
well studied clusters such as the Pleiades.  To do this,
we performed a differential iron abundance analysis with
respect to the Sun, using lines for which equivalent widths
have been published for Pleiades stars.  We analyzed our
measured widths in NGC~2516 and those in the Pleiades using
the same model atmospheres and adopted temperature scale.

We chose to work with a set of six isolated, unblended
\ion{Fe}{1} lines concentrated around 6700 \angs
(specifically 6677 \AA, 6703 \AA, 6705 \AA, 6726 \AA, 6750
\AA, and 6752 \AA).  Equivalent widths for these lines
in a set of Pleiades F dwarfs are given by \citet{bbb88}
and by \citet{bbr88}.  We found that many of the other
lines employed by Boesgaard \& Friel, specifically those
with $\lambda > 7000$ \AA, were blends, and that our
measurements of line strengths in twilight solar spectra
were not in agreement with the values quoted by Boesgaard
\& Friel.

We measured the equivalent widths of our chosen lines in
both of the two targets and in a twilight solar spectrum
using the same instrumentation and reduced in a similar
way.  We compared the solar equivalent widths with those
from the Kitt Peak Solar Atlas, finding good agreement in
all cases.  We then used the one-dimensional homogeneous
LTE model {\sc ATLAS9} model atmospheres \citep{kur93}
to calculate the iron abundances.  The atmospheres
incorporated the mixing length treatment of convection with
$\alpha = 1.25$.  We did this in a differential way with
respect to the Sun, assuming solar parameters of $T_{\rm
eff} = 5777$ K, $\log g = 4.44$, and a mictoturbulent
velocity of 1.0\kms.  The $gf$ factors for the lines were
tuned to give a solar iron abundance of 7.54 \citep{mil94}
on the usual logarithmic scale where the abundance of
hydrogen is 12.

The equivalent widths for our target stars and for the
Pleiades F stars were then fed into the same models with
$T_{\rm eff}$ determined from $B - V$ via the \citet{sh85}
relationship, assuming a solar metal abundance and
$\log g = 4.5$.  We assumed $E(B - V) = 0.12$ for
NGC~2516 \citep{lw84,dk89} and $E(B - V) = 0.04$ for the
Pleiades \citep{cp76,sh87} to obtain intrinsic colors.
We allowed the microturbulence to be a fitting parameter,
requiring that the derived iron abundance was independent
of line equivalent width.  We found that in all cases the
microturbulence was in the range 1.0 to 1.75 \kmss with an
uncertainty of about 0.25 \kms.  To investigate systematic
errors, we re-determined abundances using different
values of $T_{\rm eff}$, $\log g$, and microturbulence.
A temperature variation of $\pm 100$ K, corresponding
to a change in intrinsic color of $\mp 0.03$ mag, led to
$\Delta {\rm [Fe/H]} = \pm 0.07$; a change in $\log g =
\pm 0.2$ gave $\Delta {\rm [Fe/H]} = \pm 0.03$; a change
in the microturbulence of +0.25 \kmss led to $\Delta {\rm
[Fe/H]} = \mp 0.04$.

We find [Fe/H] $= -0.02 \pm 0.03$ for DK~320 and [Fe/H] $=
+0.13 \pm 0.04$ for DK~325, where the error is the standard
error in the mean from the six lines.  To these errors we
add about 0.08 dex to account for errors in the photometry,
gravity, and microturbulence.  The mean for NGC~2516 from
the two stars is therefore [Fe/H] = $+0.05 \pm 0.06$.
To put this onto an absolute abundance scale we should then
also have to consider errors in the assumed temperature
scale, atmospheric models, and assumed mean reddening.
We should also consider the consequences of using [Fe/H] =
0 to calculate the temperatures from the intrinsic colors.
This is in fact a small correction, because each 0.1
dex decrease in metallicity only decreases the assumed
effective temperature by $\approx 30$ K.  Thus if we had
assumed initially that [Fe/H] $ = -0.3$ for the purpose of
calculating $T_{\rm eff}$, we would have derived [Fe/H] $
= -0.01$ from the spectra, which is clearly inconsistent
with the original assumption.

There were six Pleiades stars for which all six \ion{Fe}{1}
lines had equivalent width measurements.  We find a
mean [Fe/H] $= 0.01 \pm 0.03$, assuming intrinsic color
errors of $\pm 0.02$ and assuming similar gravity and
microturbulence errors to the NGC~2516 stars.  This is
in reasonable agreement with [Fe/H] $= -0.034 \pm 0.024$
found by \citet{bf90}, who also employed the \citet{sh85}
temperature scale in their analysis.

For the purposes of comparison, the ratio of iron abundance
in NGC~2516 and the Pleiades does not depend on the adopted
temperature scale nor on the model atmospheres employed,
because both sets of data were treated in the same
way.\footnote{For example, an alternative calibration
of the color-temperature relation is provided by
\citet{hbs00}, who provide polynomial fitting functions in
both $B - V$ and $V - I$.  For the two stars in question,
this temperature scale is 66~K cooler in $B - V$ that
that from \citet{sh85} used in the abundance analysis.
For the entire sample the mean difference is $\leq 25$ K
for any reddening $E(B - V) \leq 0.2$; it just so happens
that our two stars with high S/N spectra have colors
where the temperature difference between the two $B -
V$ scales is a maximum.} The logarithmic ratio of the
iron abundances in NGC~2516 to the Pleiades is therefore
$\Delta{\rm [Fe/H]} = +0.04 \pm 0.07$, and so we 
therefore conclude that the
metallicity of NGC~2516 is roughly solar abundance.  For our
final value, we combine this differential
measurement of [Fe/H] and the Boesgaard \& Friel scale
for the Pleiades and conclude that the metallicity 
of NGC~2516 is ${\rm [Fe/H]} = +0.01 \pm 0.07$.

The only additional error is about $\pm 0.02$ in [Fe/H]
for every $\pm 0.01$ change in the relative reddening
values for each cluster. We discuss the reddening in
further detail in the next section when we look at the
CMD for NGC~2516, but for now we note that if we took an
extreme reddening for our two stars of $E(B - V) = 0.19$
\citep{sfd98}, we would derive $\Delta{\rm [Fe/H]} = +0.15$
with respect to the Pleiades.  We adopt an error in the
exction of $\pm 0.02$ so the error in our spectroscopic
determination of [Fe/H] rises to $\pm 0.08$.
Nevertheless we caution
the reader that this result has been derived from only two
stars, and there is a great need to increase this sample.
Unfortunately, most of the F stars have spectra which
have large rotational broadening, so it may be necessary
to pursue this work on the fainter G stars with larger
telescopes.

Our new abundance is consistent with at least some other
determinations of the abundance of NGC~2516.  \citet{tat97}
give [Fe/H] $= +0.060 \pm 0.030$ from DDO photometry of
two stars.  \citet{niss88} reports $uvby-\beta$ photometry
of 6 stars in NGC~2516, deriving [Fe/H] $= +0.06 \pm 0.06$
and $E(b-y) = 0.081 \pm 0.016$, which corresponds to $E(B -
V) = 0.11 \pm 0.02$.  The Nissen et al.\ result, however,
disagrees with that of \citet{lw84}, who also report
$uvby-\beta$ photometry in NGC~2516 and derive [Fe/H] $=
-0.28$ using about the same extinction as \citet{niss88}.

\subsection{The distance and reddening to NGC~2516 and
its photometric metallicity}

We now present a new determination of the cluster
distance and photometric metallicity using an updated
version of the main-sequence fitting method outlined
in \citet{pin98}.  The full details of the method may be
found in \citet{pin02}, and a preliminary determination
for NGC~2516 is discussed in \citet{pin00}.

The process of isochrone fitting separately in $B - V$
and $V - I$ yields a photometric metallicity for a cluster
since the luminosity of the main sequence at fixed color
changes more rapidly in $B - V$ (where ${\Delta M_V} /
{\Delta {\rm [Fe/H]}} \approx 1.3$) than it does in $V -
I$ (${\Delta M_V} / {\Delta {\rm [Fe/H]}} \approx 0.6$).
The signature of a metal poor cluster would then be a
distance modulus derived in $B - V$ that was greater than
found in $V - I$. Indeed, \citet{jjt98} used this technique
to derive [Fe/H] $=-0.18 \pm 0.05$ for NGC~2516, using
the smaller sample of radial velocity members available
then, while \citet{pin00} found [Fe/H] $= -0.26$ using the
same data; both studies used $E(B - V) = 0.11$. As these
are both at variance with our spectroscopic abundance
(above), we wish to explore the distance and metallicity
determinations in some detail.

Our approach here is similar in approach to that employed by
\citet{jth01} in the determination of candidate photometric
members of NGC~2516.  In that study, they generated
empirical isochrones starting with the \citet{sei00}
models, and then calibrated the isochrones with multicolor
photometry in the Pleiades, assuming a Pleiades distance
modulus of 5.6.  They derived a distance to NGC~2516
by fitting a 150 Myr isochrone to the main-sequence
appearing in $B - V$ and $V - I$ CMDs, which is relatively
distinct from the background except at faint magnitudes.
The result of this calculation determined the relative
distance between the Pleiades and NGC~2516.   The result
depended on the metallicity: for solar abundance, the
distance is $8.10 \pm 0.05$ in both $B - V$ and $V - I$,
while for half solar the distances in $B - V$ and $V - I$
were respectively $7.85 \pm 0.05$ and $7.90 \pm 0.05$.
Both calculations assumed $E(B - V) = 0.12$, and a
particular choice of the extinction law.  By comparison,
the Hipparcos parallax translates to a distance modulus of
$(m - M)_0 = 7.70 \pm 0.16$ \citep{rob99}, which is close
to the value of $7.77 \pm 0.11$ found by \citet{sung02}.
This last study presented $UBVI$ photometry for NGC~2516
independently of \citet{jth01}, and derived a photometric
metal abundance of ${\rm [Fe/H]} = -0.10 \pm 0.04$ by
comparison of the main sequence to uncalibrated isochrones
from the Padua group.  We will have more to say about the
\citet{sung02} analysis below.

\begin{figure}\epsscale{0.75}
\plotone{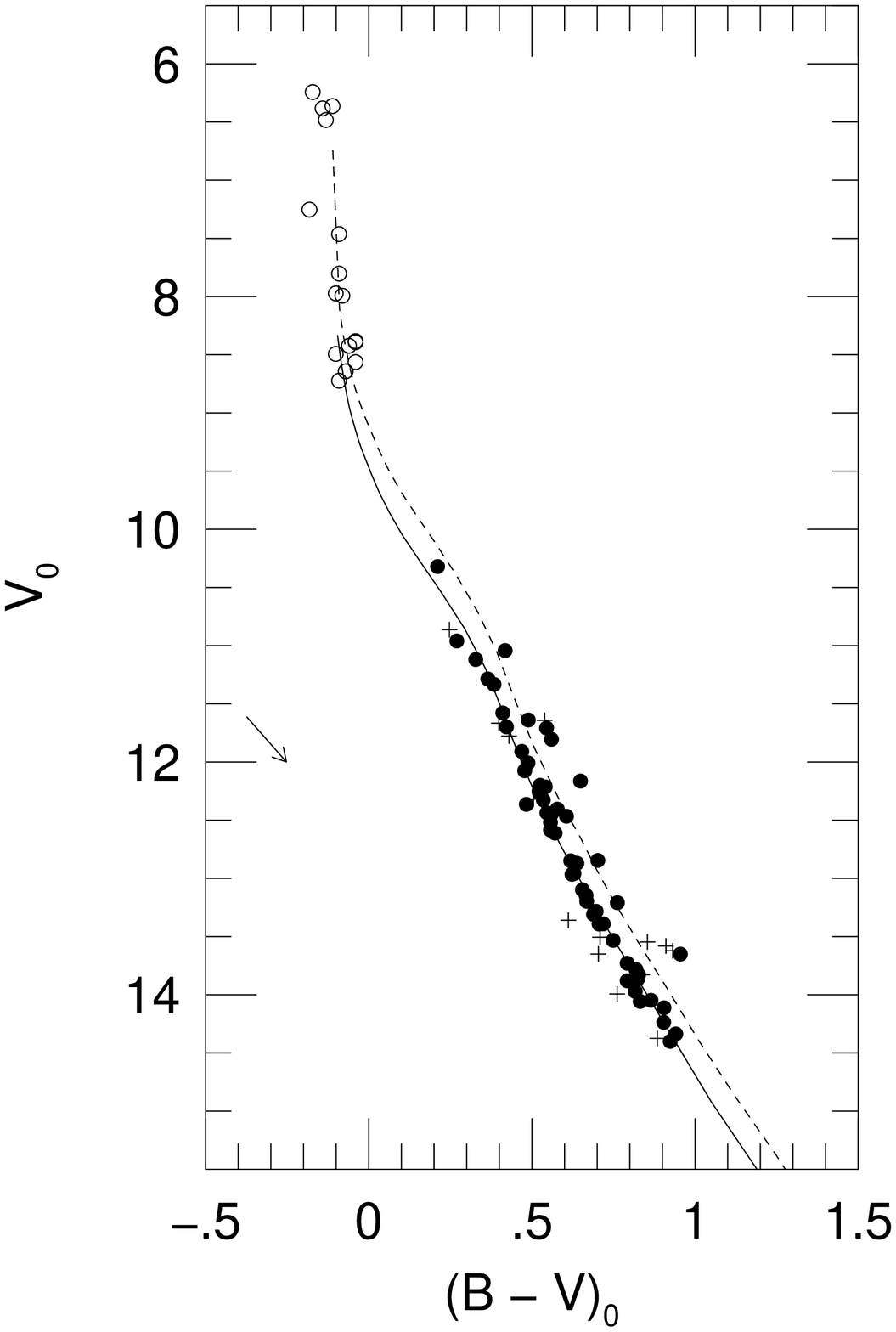} \caption{Color-magnitude diagram in
$V$, $B - V$ for NGC~2516.  Filled points are for radial
velocity members from this survey, while the open points
are for bright members in the study of \citet{gl00}.
Plusses denote non-members from this paper.  The photometry
has been dereddened using $E_0(B - V) = 0.12$ and the
extinction law derived by \citet{bcp98}.  The dashed
line shows a 140 Myr isochrone, empirically calibrated as
described in the text, shifted to $V_0 - M_V = 7.70$, the
Hipparcos distance to NGC~2516 \citep{rob99}.  The solid
line shows the solution from this paper, namely $(V_0 -
M_V) = 7.93$ and ${\rm [Fe/H]} = -0.05$.  The arrow denotes
the reddening vector.  \label{bvcmdfig}} \end{figure}

\begin{figure}\epsscale{0.75}
\plotone{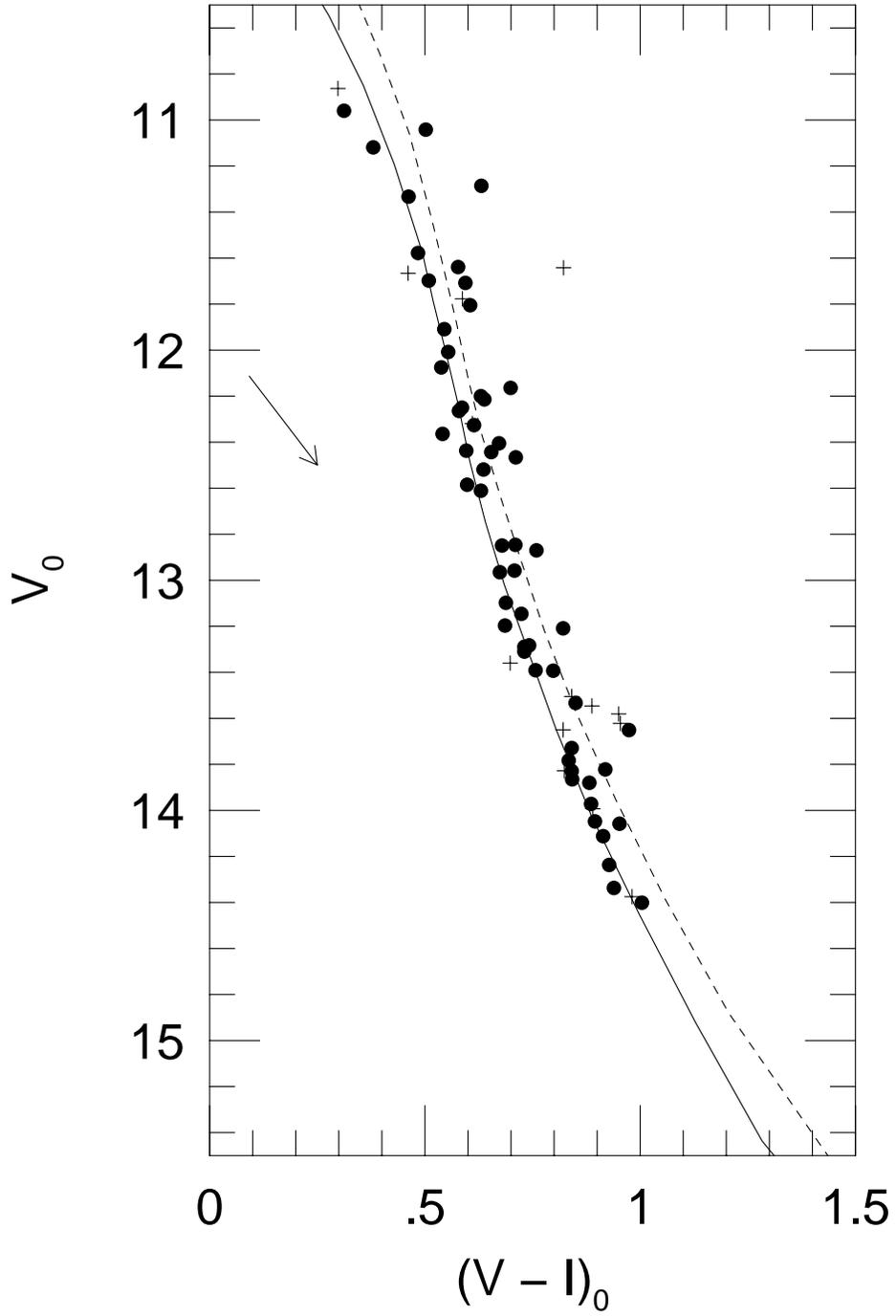} \caption{Color-magnitude diagram in
$V$, $V - I$ for NGC~2516.  Symbols and isochrones are the
same as in Fig.\ \ref{bvcmdfig}.  Note the change in scale
with respect to Fig.\ \ref{bvcmdfig}.\label{vicmdfig}}
\end{figure}

We begin the discussion by comparing the photometry of
our radial velocity members to isochrones at the distance
measured by Hipparcos.  Figure \ref{bvcmdfig} shows the
color-magnitude diagram in $V_0$ and $(B - V)_0$ for the
combined sample, where the filled points display photometry
for our radial-velocity members and the open points show
the \citet{dk89} photometry for main-sequence members
from the radial velocity study of \citet{gl00}.  The stars
deemed non-members from our radial velocities are displayed
as plusses.  Figure \ref{vicmdfig} displays a CMD in $V_0$,
$(V - I)_0$ for the magnitude range containing our sample
only. The photometry has been dereddened according the
prescription of \citet[][their Appendix F]{bcp98} using a
value $E(B - V)_0 = 0.12$ for hot stars and including the
dependence of extinction and reddening on stellar color.
In $B - V$, there is a tight single-star sequence with
a distinctly brighter binary sequence;  in $V - I$ there
is no such clear separation and the main-sequence has a
fairly wide range of colors.  The dashed lines on Figures
\ref{bvcmdfig} and \ref{vicmdfig} show an isochrone for
${\rm [Fe/H]} = 0.0$, calibrated as described below,
and relocated to the Hipparcos distance.  We conclude
from Figures \ref{bvcmdfig} and \ref{vicmdfig} that the
Hipparcos distance and the similar \citet{sung02} value
are correct only if NGC~2516 is metal poor (about ${\rm
[Fe/H]} = -0.3$), provided that the distance modulus to
the Pleiades is near 5.6.  Alternatively, NGC~2516 may
share the Pleiades ``problem'' \citep{pin98} of having
a Hipparcos parallax that makes the cluster considerably
fainter than expected for its metal abundance as derived
in this paper.

The isochrones were derived from evolutionary tracks
incorporating angular momentum transport and an updated
equation of state \citep{sil00,tpt02}.  The tracks are
first used to generate isochrones in the theoretical plane
(i.e., $M_V$, $T_{\rm eff}$).  An initial calibration
to $BVRI$ colors comes from the \citet{lcb97} model
atmospheres.  The colors are then further corrected
to match the photometry of single star members of
the Hyades\footnote{In its full detail, the 
\citet{pin02} method also includes fainter Pleiades
stars to extend the empirical calibration
to redder colors than are available for Hyades stars with
Hipparcos parallaxes; this consideration does not, however,
make a difference here to NGC~2516 since our survey only
extends to about $(B - V)_0 = 0.9$, a color range 
completely spanned by Hyades stars with excellent
Hipparcos parallaxes.} cluster;  this step is
necessary because the isochrone shapes after
initial calibration still do
not match the observed cluster main sequence at all
luminosities \citep[e.g.,][their Figs.\ 4 and 5]{ter00}.
Parallaxes for individual Hyades members, which have 
typical errors of 2\%, were used to correct the photometry
to the distance of the dynamical center of the cluster
which has a distance modulus $3.34 \pm 0.01$, or $d = 46.6
\pm 0.2$ pc \citep{per98}.  The empirical corrections 
were generated for a theoretical isochrone with [Fe/H] $= +0.13$
\citep{bf90}, with scaled solar abundances and the solar
helium abundance.  The empirical color
corrections can then be applied to isochrones of any age
and metallicity.  For metallicities within a few tenths
dex of solar the empirical corrections to the colors
are assumed to be independent of metallicity;  this is
equivalent to assuming that the effect of line blanketing
on the colors is correct in the model atmospheres employed
in the initial calibration of the isochrones.

In the \citet{pin02} method, cluster distances are then
found by determining $V_0 - M_V$ for each star, where $M_V$
is the absolute magnitude of the empirically calibrated
isochrone at the dereddened color of the star, and $V_0$
is the dereddened magnitude.  The cluster distance is found
separately in $B - V$ and $V - I$ from the median value
of $V - M_V$ in color bins.  Stars that are more than 0.1
mag from the median are rejected, which reduces the effect
of binaries on the distance determination.  The distance
modulus is derived for isochrones of several different
metallicities.  The cluster photometric metallicity is
taken as the metallicity of the isochrone for which the
distance moduli determined from $B - V$ and $V - I$ agree.
In addition, it is generally possible to use the narrowness
of the main sequence in $B - V$ as a measure of whether the
shape of a particular isochrone matches the main sequence
data in that color;  a shape mismatch is indicated by
finding that $\langle V_0 - M_V\rangle$ is a function
of color.

There are two problems which arise in determining the
distance and photometric metallicity of NGC~2516.  First,
the cluster has a markedly higher reddening than most
of the clusters in the original study of \citet{pin98}.
The derived photometric metallicity depends sensitively
on the adopted reddening value, the ratio of total to
selective extinction in the various photometric bands,
and whether there is differential reddening towards
the cluster.  Second, the large span in color of the main
sequence in $V - I$ (Fig. \ref{vicmdfig}) means that there
will be a wide range of statistically permitted values
of the apparent distance modulus in the ($V$, $V - I$)
CMD;  as a consequence, the photometric metallicity will
have a relatively large error compared to clusters with
narrow sequences.  The width of the main sequence in $V -
I$ may be caused by a significant binary fraction in the
cluster, any observational selection favoring binaries over
single stars, or by significant differential extinction
towards NGC~2516.

Previous studies of the extinction towards NGC~2516
have generally arrived at values near $E(B - V) = 0.11$
or $0.12$ \citep[][and references therein]{dk89,sung02}.
In contrast, the \citet{sfd98} dust-emission maps in the
direction of NGC~2516 indicate a higher average reddening
$\langle E(B - V)\rangle = 0.21$ to our spectroscopic
members, with a variation of $\Delta E(B - V) = 0.03$
towards our target stars.  That the average reddening
is higher than that obtained from the hot turnoff stars
in the cluster is not, perhaps, a surprise, since the
maps are based on the integrated infrared emission out
to infinity and the cluster is located at a distance of
about 400 pc at a galactic latitude of $b = -16^\circ$.
Also, some studies \citep{ag99,vbm01} have claimed that the
\citet{sfd98} maps overestimate the extinction by 30--50\%
when $A_V > 0.5$ (equivalently when $E(B - V) \geq 0.15$).

\begin{figure}\epsscale{0.75}
\plotone{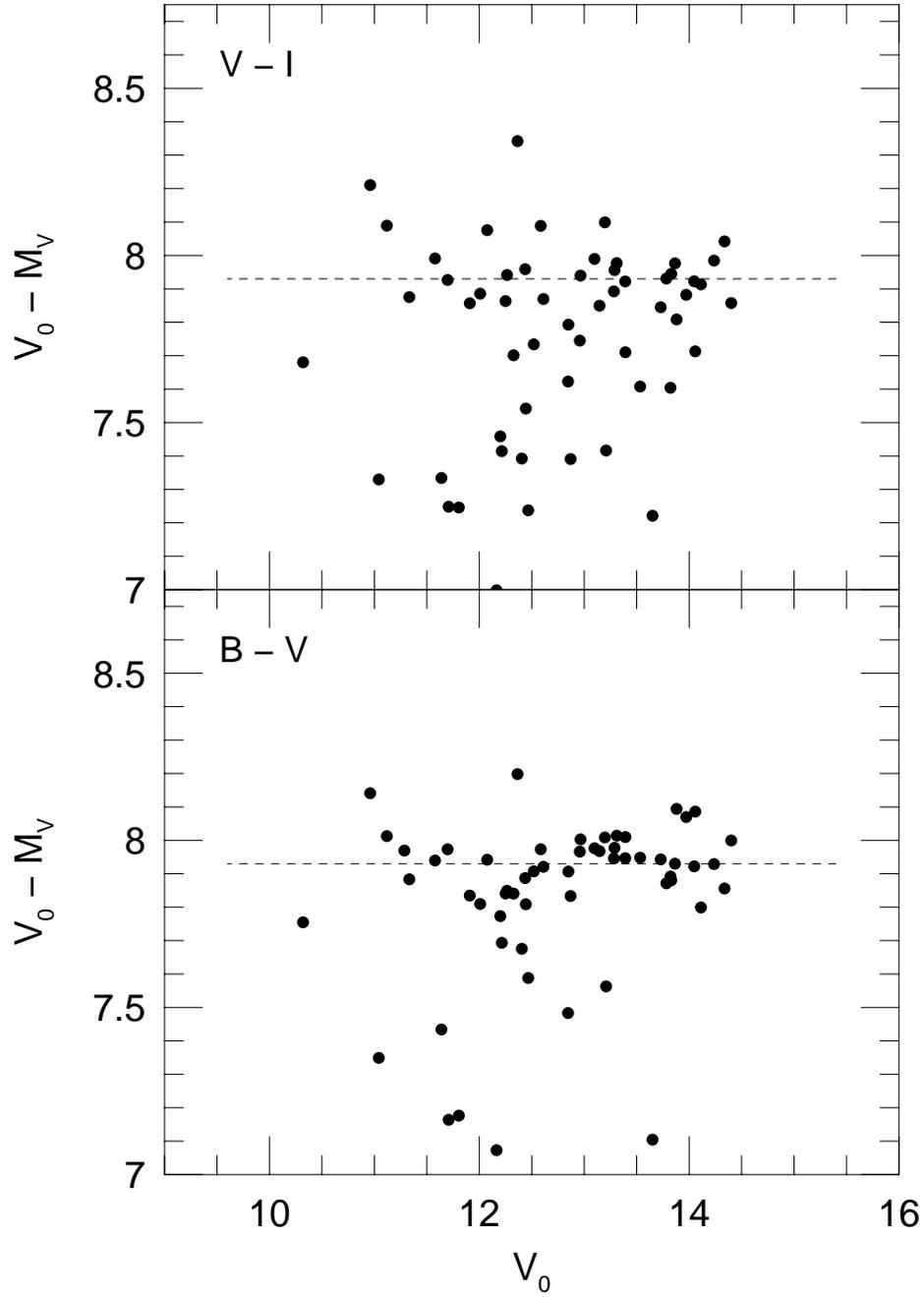} \caption{Differences between the
dereddened $V$ magnitude and an empirically calibrated
isochrone for ${\rm [Fe/H]} = -0.05$.  The top panel shows
the differences with respect to the isochrone in $V - I$,
while the lower panel shows these in $B - V$.  The dashed
line shows the best fitting distance modulus $V_0 - M_V =
7.93$.\label{distfitfig}} \end{figure}

In this analysis, we attempted to fit isochrones of
a wide variety of metallicities using the sample of
radial velocity members in Table 1, and allowed the
reddening to vary from the previously determined values.
Low metallicities (${\rm [Fe/H]} \leq -0.4$) provided
a poor fit to the isochrone shapes at any value of the
reddening.  Statistically acceptable fits were obtained
for metallicities between [Fe/H] $= -0.3$ and $+0.1$
at reddenings between $E(B - V) = 0.12$ and $0.18$.
For the canonical reddening of $E(B - V) = 0.12$, the
photometric metallicity for NGC~2516 is ${\rm [Fe/H]} =
-0.05$ at a distance of $V_0 - M_V = 7.93$.  We illustrate
this solution in Figures \ref{bvcmdfig} and \ref{vicmdfig},
where the best fitting isochrone is displayed as a solid
lines in each figure.  In $B - V$, this isochrone runs
through the single-star sequence, and runs along the lower
edge of the main sequence in $V - I$.  The match of the
isochrone shapes to the photometry is shown in Figure
\ref{distfitfig}, which shows the individual values in
$V_0 - M_V$ in each color.  There is no significant trend
with magnitude in the residuals in either color.

We extensively explored the sensitivity of the
determination of the distance and photometric metallicity
to choices of reddening and the reddening law.  The results
are summarized in Table 3.  The smallest source of error
is the accuracy of finding the median apparent distance
modulus.  The largest source of error is the value of the
extinction itself.  Adopting higher extinctions increases
the photometric metallicity appreciably but reduces
the derived distance only slightly;  the effect on the
distance modulus is smaller since the slope of reddening
vector is only somewhat different from the average slope
of the main sequence.  An equally important source of
error is the steepness of the adopted extinction law,
which may be characterized by the value of $R_{\rm VI}
\equiv E(V - I) / E(B - V)$.  Increasing $R_{\rm VI}$
results in a smaller distance and a significantly higher
photometric metallicity.  The final values we derive for
NGC~2516 are ${\rm [Fe/H]} = -0.05 \pm 0.14$ and $V_0 -
M_V = 7.93 \mp 0.14$.

Adopting an $E(B-V)$ error of $\pm 0.02$, then the
photometric metallicity we found for NGC~2516 is
consistent, within the errors, with the value of $+0.01
\pm 0.08$ we find from our spectroscopic analysis.
The main reason we derive a higher metallicity from the
CMD here than did \citet{pin00} is because we employ the
\citet{bcp98} extinction law, for which $R_{\rm VI} = 1.34$
at the mean $V - I$ of our sample, in contrast to the value
$R_{\rm VI} = 1.25$ used in the previous determinations
of the metallicity from the CMD.  In our analysis, we took
$\pm 0.04$ for the error in $R_{\rm VI}$, representing the
difference between the adopted value and the \citet{dwc78}
value of $R_{\rm VI} = 1.30$ for the average color in
our sample.  The reddening law we used here has $A(V) /
E(V - I) = 2.47$, on average, which is equal to the value
of $2.49 \pm 0.02$ determined empirically from observations
of stars in the bulge by \citet{stan96}.

The distance we determine is $1.5\sigma$ higher than the
Hipparcos value, which is not a statistically significant
difference by itself.  This difference corresponds
to a difference in parallax of 0.29 mas, quite a bit
smaller than the $1-2$ mas systematic errors in the
parallaxes needed to explain the difference between the
Hipparcos and isochrone-fitting distance to the Pleiades
\citep{pin98,ng99}.  Reconciling the photometric distance
with the Hipparcos distance would require the photometric
metallicity to be lower than we determine, which would make
that metallicity inconsistent with our spectroscopic value.

If instead we adopted our spectroscopic value and fit a
solar abundance isochrone to the narrow $B - V$ sequence
in Figure \ref{bvcmdfig}, we would obtain a more precise
distance of $(V_0 - M_V) = 8.05 \pm 0.11$, which is $2.5
\sigma$ higher than the Hipparcos distance.

Recently, \citet{sung02} analyzed new $UBVI$ photometry
of NGC~2516, and used the Padua group isochrones to
derive a photometric metallicity of ${\rm [Fe/H]} =
-0.10 \pm 0.04$ and a distance modulus $V_0 - M_V = 7.77
\pm 0.11$. Therefore despite finding a relatively high
abundance for NGC~2516, their distance is consistent with
the Hipparcos value.  In our opinion, their analysis is
seriously flawed, even though their distance determination
is marginally equal to ours within the errors.  Here we
have demonstrated that one can achieve an excellent fit
to the CMD in both $B - V$ and $V - I$ from isochrones
calibrated on the Hyades and the Pleiades.  An inspection
of Figure 2 in \citet{sung02} shows that the isochrones
they adopted do not simultaneously fit the main sequences
in $B - V$ and $V - I$ and are a poor match to the
shape of the main sequence in $B - V$ or $U - B$.  Their
determination of the distance rests primarily on fitting
an isochrone to the $(V, V - I)$ CMD; the corresponding
isochrone in $B - V$ is significantly too bright (or red)
and misses the bulk of the main sequence over at least 5
mag in $V$.  They then attribute the difference between
the isochrone and the stellar locus
to a UV excess, and claim that this
is a general feature of young stellar clusters.  Finally,
they determine the cluster metallicity by rejecting most
of the stars in the cluster as having a UV excess, which
necessarily means that they derived a value near solar
because their sample is contaminated with nonmembers.
In a separate paper \citep{sta02}, we present evidence
for a blue excess in young, low-mass stars, mainly from
data for late-K dwarfs in the Pleiades.  
We find no evidence for blue excesses
in G and early K dwarfs, and hence we do not believe
that this phenomenon should affect the results we have
presented for NGC~2516.

\subsection{Other metallicity estimates}

\citet{dn01} have recently published a detailed study of
the eclipsing binary system V392 Car, which is a member of
NGC~2516.  They compared their derived luminosity for the
members of this system against stellar structure models
using a distance close to the Hipparcos measurement,
and derived a metallicity of [Fe/H] $= 0.0 \pm 0.1$.
We performed a similar analysis, instead comparing the
stars' measured radii to that in the isochrones we used
for the distance analysis. The result is [Fe/H] $=-0.05
\pm 0.05$ for these stars by the constraint that their
radii match the isochrone radii, consistent within the
errors with our spectroscopic determination.

\section{Stellar rotation in NGC~2516}

Following the approach of \citet{sod01} and \citet{sod93},
we display in Figure \ref{vsinifig} the correlation between
\vsinis and color for NGC~2516 and other well-studied
open clusters.   From top to bottom are shown the Pleiades,
NGC~2516, M~34 and the Hyades (600 Myr).  The data sources
are as follows:  Pleiades: \citet{que98}, supplemented
with new observations in \citet{ter00}; NGC~2516:
this paper; M~34: \citet{sod01}; Hyades: $v \sin i$
data from \citet{kra65,kra67}, and rotational periods
from \citet{rad87} converted to $v \sin i$.  The solid
line in each panel is a representation of the Hyades data,
while the dashed line indicates the location of the rapid
rotators in the Pleiades.  For NGC~2516, we adopted $E(B -
V) = 0.12$ (above).

In Fig.\ \ref{vsinifig}, the time evolution of angular
momentum from the Pleiades to the Hyades (100 to 600 Myr)
manifests itself as a gradual reduction in \vsinis for the
slow rotators, and a significant decline in the numbers of
rapid rotators.  NGC~2516, like the Pleiades, has a number
of rapid rotators, and these have rotation speeds about
70\% of the stars in the Pleiades with similar colors.
The number of rapid rotators declines markedly by the
age of M~34, which has an age estimated from 180 Myr
\citep{mmm93} to $\sim 250$ Myr \citep{jp96}.  In M~34,
the upper bound of rotation rates is about a factor of two
lower than it is in the Pleiades.  Our observations in
NGC~2516 do not extend to the faint magnitudes at which
the K dwarfs in the Pleiades and M~34 are still rapidly
rotating.

In Figure \ref{models}, we compare the rotation rates
of stars in the Pleiades, NGC 2516, M~34 and the Hyades
with theoretical models. The data sources are the same as
in Figure \ref{vsinifig}. The conversion between color
and effective temperature is taken from \citet{hbs00}.
The models are taken from \citet{sil00}, and are for solar
metallicity stars.  The adopted ages of the clusters are
shown in figure \ref{models}, where we have taken the
\citet{jp96} age of 250 Myr for M~34 as in their paper.
In the models the stars are assumed to be differentially
rotating. The loss of angular momentum from the surface
was modelled using a saturated magnetic wind loss law
\citep{cdp95} with a mass-dependent saturation threshold
$\omega_{\rm crit}$ which follows a Rossby scaling down to
$M = 0.5 M_{\odot}$ and then chosen to match the low mass
rapid rotators in the Hyades.  We used five different
values of the disk-locking lifetime, during which the
star is assumed to be rotating at the same rate as its
proto-stellar disk. The lifetimes are chosen to span the
expected range of proto-stellar disk lifetimes, up to
10 Myr.

The overall impression from Figure \ref{models} is that the
rotation rates of the stars in NGC 2516 are similar to what
would be expected at an age near 140 Myr if the angular
momentum evolution of intermediate mass stars ($M = 0.6 -
1.1 M_{\odot}$) in all clusters was similar.  In this plot,
M~34 has a few stars with rotation rates significantly
higher than the fastest-spinning model.  This may represent
a departure from the suggested uniformity of the Pleiades :
NGC~2516 : Hyades evolution, or else the age of M~34 is
closer to the value of 180 Myr from \citet{mmm93} than
the 250 Myr shown here.  Establishing that different star
clusters represent time snapshots of a universal evolution
of angular momentum thus would require better age estimates
for these clusters than is currently available.

The data for both NGC 2516 and M~34 probe only the
intermediate mass stars, and do not extend to the low mass
stars where the standard Rossby scaling breaks down. There
is a substantial change in behavior of the low mass models
in this age range depending on the choice of the saturation
threshold (standard Rossby scaling vs.\ the \citet{sil00}
fit to the Hyades data).

If initial conditions for star formation were universal,
we would expect that the distribution of disk lifetimes
would be the same from cluster to cluster, and should
not depend on the age of the cluster. It appears that
the Pleiades and NGC 2516 have similar fractions of
stars with disk lifetimes between 0 and 1 Myr. However,
the expected rotation rates for stars with disk lifetimes
longer than 1 Myr are slow enough that they overlap with
our observational limits. We cannot say whether the
distribution is the same without better resolution of
rotation rates or photometric period determinations.

In Figure \ref{xrayfig} we plot the fractional luminosity
in X-rays ($L_{\rm X} / L_{\rm bol}$) for NGC~2516 against
the inverse of Rossby number, which is defined
as the ratio of the rotation period to the
turnover of time of a convective cell at the
base of the convective zone.  The X-ray
luminosities were taken from \citet{jtp97}; filled points
show measured values, while filled triangles show upper
limits on the X-ray detection.  The open points show
equivalent data for a set of Pleiades stars with
known rotation periods \citep{kri98}.  
In order to treat both data sets the same way, we
ignored the rotation periods for the Pleiades stars
and instead began with \vsinis values for these.
The measured values for \vsinis were used to
compute a rotation period using the empirically calibrated
isochrones employed above, and the isochrones were used
to generate relations between effective temperature and
the $B - V$ and $V - I$ colors.  The convective overturn
time was computed from the effective temperature using
the prescription in \citet{kd96}. We did not apply
a statistical correction for projection to the data,
which would have increased the \vsinis values by a factor of
$4 / \pi = 1.27$. The conclusion from
Figure \ref{xrayfig} is that both the Pleiades and NGC~2516
have the same relation between convection zone depth and
X-ray activity.  The Pleiades stars plotted here have a
higher average rotation rate and X-ray luminosity, but the
(fewer) rapid rotators in NGC~2516 are similar.  The bulk
of the NGC~2516 data lie amongst the more slowly-rotating
Pleiades;  again there is no obvious difference between
the two clusters.

\section{Summary and discussion}

The principal result of this paper is the demonstration
that NGC~2516 has approximately solar metallicity,
which we showed from an equivalent width analysis on high
signal-to-noise spectra and from the location of the main
sequence on CMDs.  While the metallicity from the latter
method has rather large errors, both random and systematic,
it is nevertheless consistent with the metallicity found
from our spectra.  We also showed that the rotational and
X-ray properties of the F and early G stars in NGC~2516
are consistent with what would be expected at an age near
140 Myr if angular momentum evolution in different clusters
is a uniform time sequence.

Why, then, did previous studies come up with values
of the metallicity that were almost always metal poor?
Both \citet{cam85} and \citet{jtp97} concluded the cluster
was metal poor by modeling the colors in the $U - B$,
$B - V$ plane, where the data were a combination of
photographic and photoelectric photometry.  The F stars
in NGC~2516 clearly showed an ultraviolet excess compared
to similar stars in the Pleiades (\citet{jtp97}, their
Figs. 10 and 11).  The new CCD photometry that we report
here from \citet{jth01} compares quite favorably to the
previous values in $B - V$ over the magnitudes of stars
in our sample, though the new study does not include
photometry in $U$ for an independent check of the zero
point in that band.

On the other hand, our conclusion that NGC~2516 has
approximately solar metallicity helps explain many other
observations of the cluster.  For example, \citet{mic00}
analyzed Rosat HRI observations of NGC~2516 stars,
and showed that the X-ray activity in this cluster is
consistent with that observed in dK and dM stars of similar
age, contrary to the expectation that the convection
properties in NGC~2516 stars would be different than in
solar-metallicity clusters.  The depletion patterns of
Li in this cluster are consistent with a metallicity near
solar \citep{jjt98}, but would require nonstandard mixing
scenarios if the cluster were metal poor.

\begin{figure}\epsscale{0.75}
\plotone{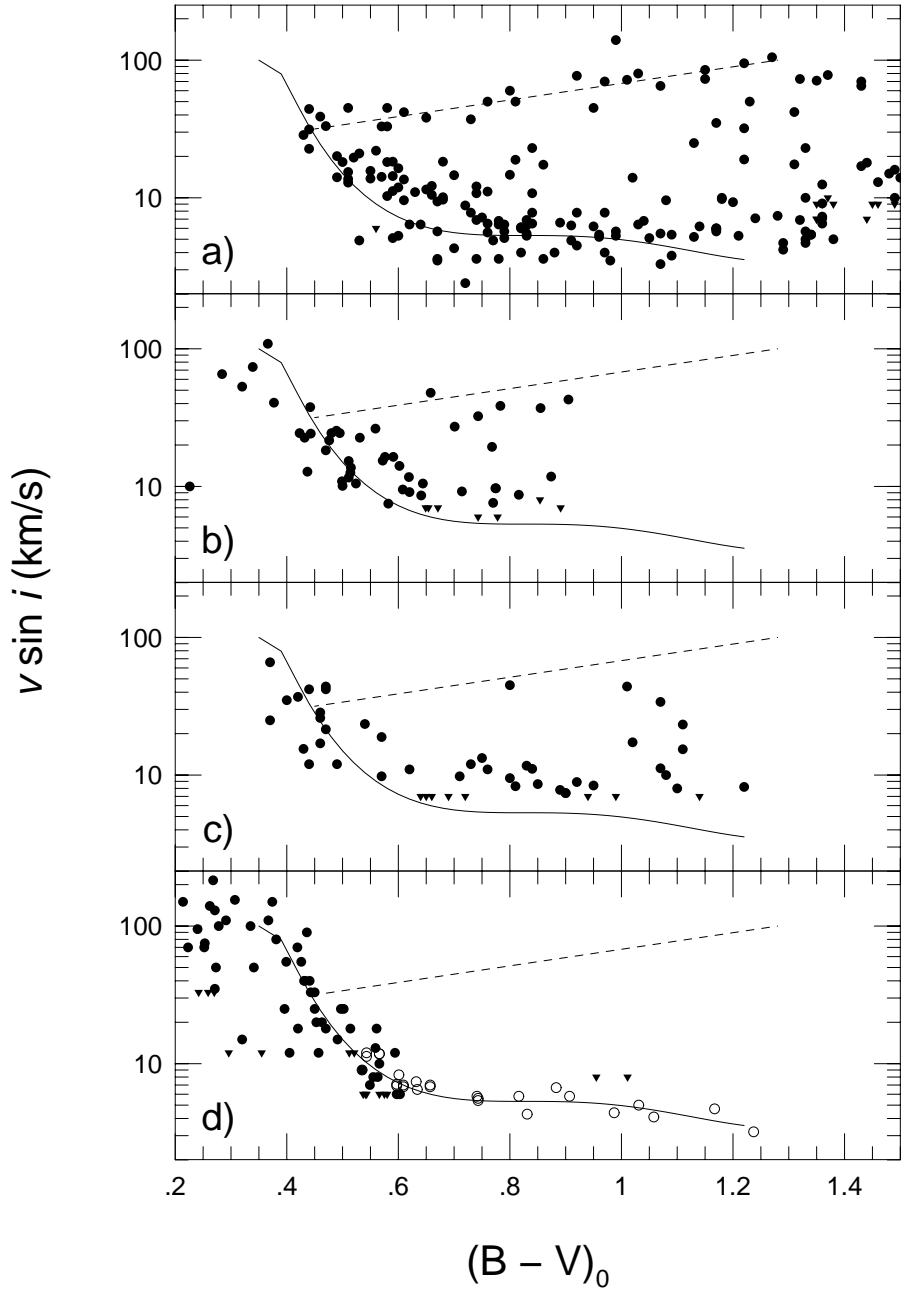} \caption{Comparison of \vsinis for (a)
Pleiades, (b) NGC~2516, (c) M~34, (d) Hyades.  Sources for
the rotation rates are described in the text.  The solid
line represents the mean Hyades relation, while the dashed
line characterizes the fast rotators in the Pleiades; these
lines are the same in all panels.  Filled points denote
measured \vsinis values, while open points show \vsinis
derived from photometric rotation periods.  Upper limits
to \vsinis are displayed as triangles. \label{vsinifig}}
\end{figure}

\begin{figure}\epsscale{1.00}
\plotone{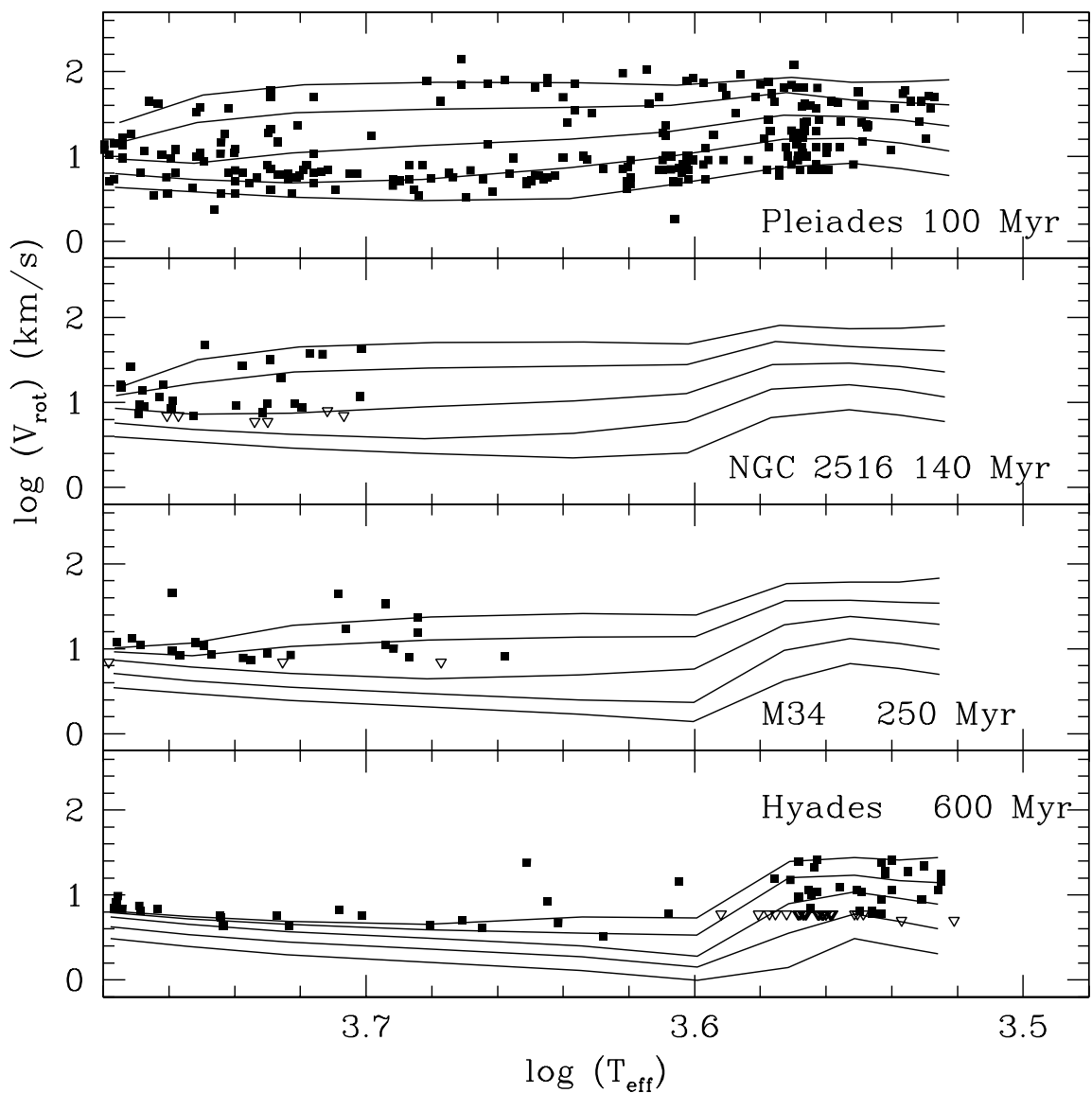} \caption{Rotation rate as a function
of effective temperature for stars in young open clusters
with ages between 100 Myr and 600 Myr. The assumed ages
of the clusters are shown in each figure. The solid
points are detections and the open triangles are upper
limits. The solid lines show the five different values of
the disk-locking lifetime that were used: 0, 0.3, 1, 3, and
10 Myr (from top to bottom). \label{models}} \end{figure}

\begin{figure}\epsscale{0.75}
\plotone{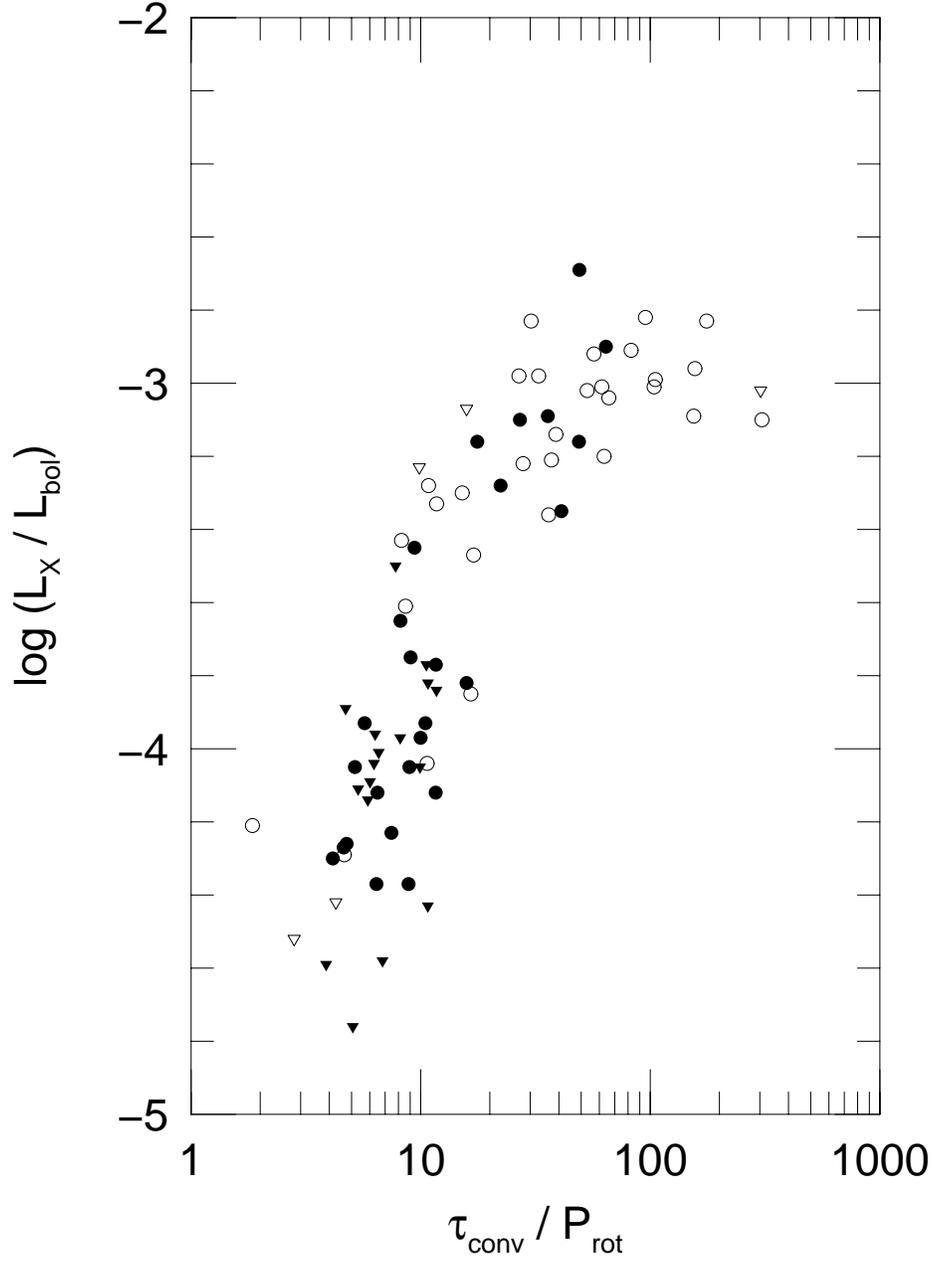} \caption{X-ray luminosity against inverse
Rossby number.  Solid points show stars in NGC~2516,
while open points are Pleiades stars with measured
periods.  Triangles show upper limits.  See text for
details.\label{xrayfig}} \end{figure}

Extending our work to lower masses where rapid rotation
is observed (e.g., Figure \ref{vsinifig}) is clearly
warranted.  Since NGC~2516 is significantly more distant
than the Pleiades, \vsinis studies amongst the M dwarfs
would require telescopes in the 8-10m range (M dwarfs have
$V \geq 14$ in the Pleiades, or $V > 16.6$ in NGC~2516).

\acknowledgements

We would like to thank
the directors and staff of both CTIO
and the Anglo-Australian Observatory.  AF was a Particle
Physics and Astronomy Research Council (PPARC) research
student during this work.  DMT and MHP acknowledge partial
support from the National Science Foundation, via grants
AST-9528227 and AST-9731621 to The Ohio State University.

\clearpage

\clearpage 

\clearpage 

\clearpage 

\clearpage 

\clearpage 

\clearpage 

\clearpage \input{table1}

\clearpage \input{table2}

\input{table3}

\end{document}

%% file: table1.tex
\begin{deluxetable}{lrrrrrrrrrl}
\tablecaption{NGC 2516 data}
\tablewidth{0pt}
\tablehead{
 \colhead{Star} & \colhead{Source} & 
 \colhead{$v(r)$} & \colhead{$\sigma(v)$} &
 \colhead{$v \sin i$} & \colhead{$\sigma$} &
 \colhead{$V$} & \colhead{$B - V$} & \colhead{$V - I$} &
 \colhead{JTH} & phot?
}  % end of table heading
\startdata
CTIO-1  & 2   & 24.2 & 1.0  & $\leq$ 8 & \nodata & 14.701 &  1.014  &  1.094  &  6049 & Y \\ 
CTIO-2  & 2   & 23.3 & 0.6  & $\leq$ 6 & \nodata & 14.184 &  0.903  &  1.007  &  6263 & Y \\  
CTIO-3  & 2   & 24.7 & 0.6  & $\leq$ 6 & \nodata & 14.287 &  0.938  &  1.007  &  6676 & Y \\ 
CTIO-5  & 2   & 24.8 & 0.8  & $\leq$ 7 & \nodata & 14.804 &  1.051  &  1.105  &  8654 & Y \\
CTIO-6  & 2   & 22.1 & 1.6  &     11.7 & 3.0     & 13.590 &  0.779  &  0.889  &  9361 & Y \\     
CTIO-7  & 2   & 23.7 & 0.7  &     10.5 & 1.1     & 13.735 &  0.804  &  0.896  & 10753 & Y \\  
DK078   & 1   & 27.4 & 0.9  &     24.2 & 2.2     & 12.068 &  0.603  &  0.741  &  7864 & Y \\  
DK081   & 2   & 26.4 & 5.0  & \nodata  & \nodata & 10.725 &  0.328  &  0.445  &  8058 & Y \\  
DK206   & 1,2 & 23.2 & 1.0  &     18.3 & 4.6     & 12.696 &  0.637  &  0.743  &  7967 & Y \\  
DK215   & 1   & 25.9 & 0.8  &     21.6 & 1.2     & 12.682 &  0.636  &  0.750  &  7104 & Y \\  
DK221   & 2   & 23.4 & 0.8  & $\leq$ 7 & \nodata & 13.293 &  0.814  &  0.875  &  6880 & Y \\  
DK230   & 1   & 27.1 & 0.5  &     12.8 & 1.0     & 12.792 &  0.597  &  0.704  &  6700 & N \\  
DK232   & 3   & 24.2 & 0.6  &     10.0 & 7.0     & 11.370 &  0.386  &  0.474  &  6446 & Y \\  
DK233   & 1   & 22.8 & 0.6  &     12.6 & 1.0     & 12.240 &  0.674  &  0.769  &  6029 & Y \\
DK234   & 1   & 20.0 & 0.7  &      9.2 & 1.2     & 13.661 &  0.874  &  0.987  &  5887 & Y \\
DK245   & 1   & 24.8 & 0.4  &      9.5 & 1.1     & 13.541 &  0.768  &  0.853  &  4914 & Y \\
DK307   & 3   & 21.2 & 2.1  &     73.8 & 1.6     & 11.753 &  0.499  &  0.625  &  7189 & Y \\
DK308   & 1,2 & 23.9 & 1.3  &     10.5 & 1.6     & 13.047 &  0.684  &  0.794  &  6452 & Y \\
DK311   & 1   & 24.0 & 3.1  &     37.7 & 3.7     & 12.437 &  0.602  &  0.718  &  5833 & Y \\
DK313   & 1,2 & 24.8 & 0.9  &     15.4 & 1.3     & 13.289 &  0.732  &  0.843  &  5862 & Y \\
DK320   & 1,2 & 24.8 & 0.8  &     13.7 & 1.1     & 12.878 &  0.675  &  0.818  &  3208 & Y \\
DK323   & 1   & 23.7 & 0.8  &     22.6 & 2.1     & 12.503 &  0.592  &  0.701  &  3621 & Y \\
DK325   & 1,2 & 23.3 & 0.5  &     11.6 & 0.8     & 13.020 &  0.671  &  0.762  &  4574 & Y \\
DK403   & 1   & 30.4 & 1.2  &     26.3 & 1.7     & 12.905 &  0.719  &  0.876  &  7585 & Y \\
DK409   & 2   & 22.5 & 0.7  & $\leq$ 7 & \nodata & 13.730 &  0.809  &  0.907  &  6852 & Y \\
DK415   & 2   & 24.8 & 1.5  &     15.3 & 2.4     & 12.954 &  0.671  &  0.800  &  5901 & Y \\
DK428   & 1   & 22.2 & 0.5  &     16.4 & 0.7     & 13.312 &  0.751  &  0.924  &  4105 & Y \\
DK463   & 2   & 25.5 & 0.6  &      7.5 & 1.0     & 13.399 &  0.742  &  0.873  &  6570 & Y \\
DK513   & 1   & 23.5 & 1.2  &     24.4 & 2.1     & 12.647 &  0.655  &  0.802  &  7743 & Y \\
DK514   & 2   & 24.1 & 0.6  & $\leq$ 7 & \nodata & 13.839 &  0.831  &  0.922  &  7312 & Y \\
DK526   & 1   & 24.9 & 1.1  &     25.4 & 1.9     & 12.759 &  0.649  &  0.778  &  6976 & Y \\
DK609   & 1   & 24.4 & 0.4  &     10.9 & 1.0     & 12.142 &  0.659  &  0.758  &  9852 & Y \\
DK612   & 3   & 25.1 & 0.5  &     65.5 & 4.7     & 11.534 &  0.444  &  0.542  & 10301 & Y \\
DK664   & 1   & 25.9 & 1.8  &     47.9 & 1.6     & 13.840 &  0.818  &  0.963  &  9465 & Y \\
DK668   & 1   & 19.6 & 1.8  &     42.8 & 1.2     & 14.120 &  1.065  &  1.141  &  8645 & Y \\
DK722   & 1   & 24.5 & 0.6  &      8.6 & 0.9     & 13.756 &  0.801  &  0.896  & 11716 & Y \\
DK742   & 3   & 22.5 & 0.7  &     53.1 & 1.0     & 11.704 &  0.480  &  0.795  & 14681 & N \\
DK803   & 2   & 23.2 & 1.1  &     16.4 & 2.2     & 13.405 &  0.736  &  0.838  &  9140 & Y \\
DK820   & 1   & 22.6 & 0.4  &      9.7 & 1.2     & 14.279 &  0.934  &  1.085  & 12118 & Y \\
DK875   & 3   & 25.6 & 1.3  &    108.8 & 8.4     & 12.000 &  0.526  &  0.647  & 12302 & Y \\
DK902   & 2   & 20.8 & 5.0  & \nodata  & \nodata & 11.465 &  0.533  &  0.665  &  8458 & Y \\
DK904   & 1   & 27.6 & 1.0  &     22.6 & 1.0     & 12.841 &  0.691  &  0.836  &  9054 & Y \\
DK907   & 3   & 25.1 & 1.0  &     24.4 & 0.5     & 12.336 &  0.583  &  0.709  &  8529 & Y \\
DK908   & 2   & 24.7 & 0.7  &     10.1 & 1.1     & 12.871 &  0.660  &  0.760  &  8536 & Y \\
DK914   & 2   & 24.5 & 0.6  &      9.1 & 1.0     & 13.641 &  0.780  &  0.851  &  9446 & Y \\
DK933   & 1   & 23.1 & 0.5  &     14.1 & 0.8     & 12.607 &  0.762  &  0.864  &  9486 & Y \\
E010    & 2   & 23.2 & 2.7  &     32.4 & 3.3     & 14.335 &  0.903  &  1.048  &  9676 & Y \\
E020    & 1   & 25.1 & 0.4  &      9.7 & 1.3     & 14.322 &  0.935  &  1.008  &  7619 & Y \\
E027    & 2   & 27.1 & 1.8  &     19.4 & 2.4     & 14.429 &  0.928  &  1.052  &  7678 & Y \\
E063    & 1   & 24.9 & 0.4  &      8.7 & 1.1     & 14.509 &  0.976  &  1.061  & 10871 & Y \\
E095    & 1   & 25.0 & 0.4  &      7.6 & 1.2     & 14.240 &  0.930  &  1.000  &  9286 & Y \\
GSC89111169  &  1,2  &  22.1  &  2.7  &  38.5  &  2.9 &  14.516  &  0.943  &  1.118  &  8634 & Y \\
GSC89111238  &  1  &  24.5  &  1.1  &  27.2  &  1.3   &  13.984  &  0.861  &  1.016  &  4598 & Y \\
GSC89111475  &  1  &  24.6  &  1.2  &  37.1  &  1.4   &  14.576  &  1.015  &  1.080  &  5957 & Y \\
GSC89113284  &  1  &  22.6  &  0.5  &  11.8  &  1.2   &  14.867  &  1.034  &  1.171  & 10040 & Y \\
JTH~758 & 3   & 23.7 & 0.7  &     24.4 & 2.0     & 12.632 &  0.640  &  0.794  &   $\cdots$ & Y \\
JTH~14223 & 3   & 23.0 & 0.7  &     40.6 & 1.4     & 12.121 &  0.537  &  0.672  & $\cdots$ & Y \\ 
\cutinhead{Nonmembers}
DK306  &  1  &  82.0  &  6.9  &  53.1  &  2.5         &  13.747  &  0.726  &  0.856  &  7147 & N \\
DK365  &  1  &  91.3  &  0.8  &  $\leq$ 7.0 & \nodata &  13.968  &  1.022  &  1.108  &  6705 & Y \\
DK417  &  1  &   9.0  &  1.1  &  17.9  &  1.4         &  12.053  &  0.515  &  0.619  &  6096 & Y \\
DK503  &  1  &  65.6  &  1.9  &  32.4  &  3.4         &  12.706  &  0.620  &  0.768  &  7650 & Y \\
DK508  &  2  &  43.6  &  2.6  & 126.3  &  5.2         &  11.250  &  0.376  &  0.456  &  6949 & Y \\
DK573  &  1  &   0.8  &  0.7  &  $\leq$ 7  &  \nodata &  14.380  &  0.874  &  1.048  &  6307 & N \\
DK574  &  1  &  77.0  &  0.6  &  $\leq$ 7  &  \nodata &  14.009  &  1.043  &  1.112  &  6724 & Y \\
DK625  &  1  & 337.0  &  0.4  &  $\leq$ 7  &  \nodata &  13.892  &  0.822  &  0.999  & 11031 & Y \\
DK669  &  1  & --0.4  &  0.5  &  $\leq$ 7  &  \nodata &  13.933  &  0.966  &  1.046  &  8689 & Y \\
DK813  &  1  &  38.3  &  0.5  &  8.8  &  0.5          &  12.029  &  0.654  &  0.980  & 10390 & Y \\
DK919  &  1  &  13.8  &  0.3  &  7.3  &  1.3          &  14.215  &  0.950  &  0.982  &  8262 & Y \\
DK967  &  2  &  56.0  &  0.5  &  10.0  &  7.0         &  12.164  &  0.546  &  0.745  & 12923 & Y \\
E013   &  1  &  31.2  &  0.5  &  9.9  &  1.3          &  14.762  &  0.996  &  1.139  &  8967 & Y \\
E111   &  1  &   5.1  &  0.5  &  $\leq$ 7.0 & \nodata &  14.037  &  0.817  &  0.979  &  6649 & Y \\
\tablecomments{Sources for the data: (1) Observations at CTIO (this study);
(2) Jeffries et al.\ (1998); (3) Observations at AAT (this study).}
\enddata
\end{deluxetable}

%% file: table2.tex
\begin{deluxetable}{lcc}
\small
\tablecaption{Cluster radial velocity}
\tablewidth{0pt}
\tablehead{
\colhead{Sample} &
\colhead{r.v. (km s$^{-1}$)} &  \colhead{$N$}
}
\startdata
All stars (this study)        & $24.2 \pm 0.2$ & 57 \\
Jeffries et al. (1998)        & $23.8 \pm 0.3$ & 24 \\
Gonz\'alez \& Lapasset (2000) & $22.0 \pm 0.2$ & 22 \\
\enddata
\end{deluxetable}

%% file: table3.tex
\begin{deluxetable}{lccc}
\small
\tablecaption{Error budget for main-sequence fit}
\tablewidth{0pt}
\tablehead{
\colhead{Source of error} & \colhead{value} & 
\colhead{$\Delta(V_0 - M_V)$} & \colhead{$\Delta({\rm [Fe/H]})$}
}
\startdata
Fit to data points       &    \nodata & $\pm 0.015$ & $\pm 0.02$ \\
Error in $E(B - V)$      & $\pm 0.02$ & $\mp 0.020$ & $\pm 0.10$ \\
Error in $E(V-I)/E(B-V)$ & $\pm 0.02$ & $\mp 0.065$ & $\pm 0.06$ \\
Total error              &    \nodata & $\mp 0.14$\tablenotemark{a}  & $\pm 0.14$ \\
\enddata
\tablenotetext{a}{Including the effect of the abundance error, which
is $\Delta(V_0 - M_V) / \Delta{\rm [Fe/H]} = +1.3$.}
\end{deluxetable}